\documentstyle[pre,multicol,aps,epsfig,amsmath]{revtex}

\begin{document}

\draft

\title{
  Kink Dynamics in a Topological $\phi^4$ Lattice
}

\author{
  A. B. Adib\footnote{e-mail: adib@fisica.ufc.br},
  C. A. S. Almeida\footnote{e-mail: carlos@fisica.ufc.br}
}

\address{
  Departamento de F\'{\i}sica, Universidade Federal do Cear\'{a},\\
  Caixa Postal 6030, 60455-760 Fortaleza, Cear\'{a}, Brazil
}

\maketitle

\begin{abstract}
  It was recently proposed a novel discretization for nonlinear Klein-Gordon field
  theories in which the resulting lattice preserves the topological (Bogomol'nyi) lower bound
  on the kink energy and, as a consequence, has no Peierls-Nabarro barrier even for large
  spatial discretizations ($h\approx1.0$). It was then suggested that these ``topological
  discrete systems'' are a natural choice for the numerical study of continuum kink dynamics.
  Giving particular emphasis to the $\phi^4$ theory, we numerically investigate kink-antikink
  scattering and breather formation in these topological lattices. Our results
  indicate that, even though these systems are quite accurate for studying {\em free} kinks in
  coarse lattices, for legitimate {\em dynamical} kink problems the accuracy is rather restricted
  to fine lattices ($h\approx0.1$). We suggest that this fact is related to the breaking of
  the Bogomol'nyi bound during the kink-antikink interaction, where the field profile loses
  its static property as required by the Bogomol'nyi argument. We conclude, therefore, that
  these lattices are {\em not} suitable for the study of more general kink dynamics, since a
  standard discretization is simpler and has effectively the same accuracy for such resolutions.
\end{abstract}



\begin{multicols}{2}

  Coherent structures in nonlinear Klein-Gordon field theories are an active topic of
research in many branches of physics, arising either as a fundamental excitation in particle
physics or as a collective excitation in effective field theories. Of particular interest
are the so-called topological solutions, where the ``kinks'' are probably among the most
studied ones (see, e.g. \cite{rajaraman} and \cite{dodd}).

  Focusing on the numerical study of kink dynamics, Speight and Ward \cite{speight1} suggested
a new discrete sine-Gordon system which preserves the so-called ``Bogomol'nyi bound'' on the
kink energy \cite{bogomolnyi}, being also applied in a $\phi^4$ system \cite{speight2} and
recently generalized to nonlinear Klein-Gordon equations in one spatial dimension
\cite{speight3}. These systems were then called ``topological discrete systems'' (TDS) by the
authors (since the Bogomol'nyi bound depends on the topological charge, i.e., on the
boundary conditions at infinity), and they showed that such lattices do not present the
so-called Peierls-Nabarro barrier, which, for continuum field theories, is a rather
artificial and undesirable effect induced by standard discretizations.

  By means of numerical experiments with the velocity of a single kink,
these authors also showed that the numerical drift in a TDS is significantly reduced
even for large spatial discretizations $h\approx1.0$, and concluded that this construction
provides an efficient and natural choice for numerically simulating kink dynamics in
nonlinear Klein-Gordon models. If the latter statement were indeed true, it is clear that
we would benefit a lot in both computational time and memory usage by using coarse
lattices with $h\approx1.0$ when studying complex kink behavior
in these field theories, such as breather formation and multiple kink (antikink) scattering.

  However, and for our disappointment, we found that a TDS choice for such dynamical problems
is good only for fine spatial discretizations $h\approx0.1$, where a simpler standard
discretization is also accurate. Our proposal in this report is to show how we have
investigated these facts.

  The equation of motion for our problem is the well-known Klein-Gordon equation:

\begin{equation} \label{eqmotion}
  \partial_{tt}\phi = \partial_{xx}\phi - V'(\phi),
\end{equation}
with subscripts denoting partial derivatives and primes denoting derivatives with respect
to $\phi$. The potential we have adopted is the (double-well) $\phi^4$ one:

\begin{equation} \label{phi4pot}
  V(\phi) = \frac{1}{4}\left( 1 - \phi^{2} \right)^{2}.
\end{equation}
Notice that it differs from that of Speight \cite{speight2} by a mere factor of 2. We have, of
course, corrected the TDS equations in order to account for this fact, and a test of validity of
this rather simple correction will be given below for the case of a free kink
(All physical quantities in this work will be given in dimensionless units. For specific
applications, a simple rescaling of the variables in the above equations can
introduce the desired units).

  In order to integrate Eq. (\ref{eqmotion}) we use the standard staggered leapfrog scheme,
which ensures second-order accuracy in time (see, e.g. \cite{recipes}):

\begin{eqnarray} \label{leapfrog}
  \dot{\phi}^{n+1/2}_{i} & = & \dot{\phi}^{n-1/2}_{i} + \Delta t [ \partial_{xx} \phi^{n}_{i} - V'(\phi^{n}_{i}) ] \nonumber \\
  \phi^{n+1}_{i} & = & \phi^{n}_{i} + \Delta t \dot{\phi}^{n+1/2}_{i}
\end{eqnarray}
where superscripts (subscripts) denote temporal (spatial) indices, overdots represent
partial time derivatives and primes are derivatives with respect to $\phi$. The spatial derivative
is evaluated with a standard second-order centered difference
[$\partial_{xx} \phi_i \approx h^{-2} (\phi_{i+1}-2\phi_{i}+\phi_{i-1})$] which coincides with
the resulting discretization required by our particular TDS choice \cite{speight2}. Being a symplectic
algorithm, the leapfrog scheme is suitable to Hamiltonian systems (see e.g. \cite{adiabdamp})
and therefore we expect good energy conservation throughout our simulations. Indeed, in a typical
simulation with $h=0.1$, the energy is conserved to better than one part in $10^{3}$.

\begin{figure}
  \centerline{ \epsfig{figure=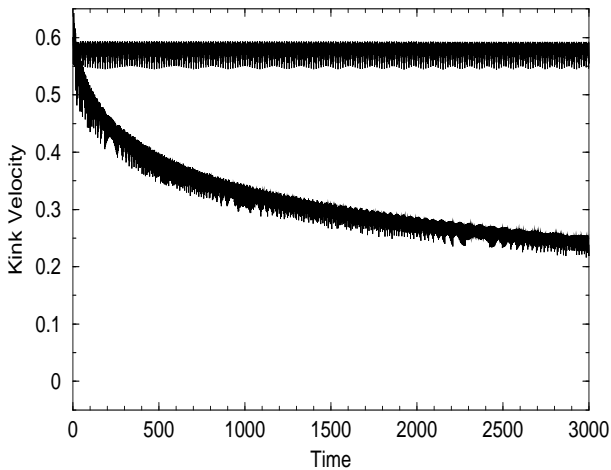,width=230pt,height=180pt} }
  \caption[caption]{
    Time evolution of the free kink velocity for $h=1.2$. The upper curve is for the TDS lattice
    whereas the lower one is for a standard discretization. The initial velocity of the
    Lorentz boosted kink is $v=0.6$.
  }
  \label{FIG:reprod}
\end{figure}

  The ``standard'' discretization of the potential (\ref{phi4pot}) is straightforward, and
involves simply the replacement of the continuum arguments $x$ and $t$ by their discrete
counterparts, e.g. $i$ and $n$. Put briefly, the TDS approach adopts a rather different
lattice potential, which for the $\phi^4$ case was suggested to have the form \cite{speight2}:

\begin{equation} \label{tdspot}
  V_{\text{TDS}}(\phi_{i}) = \frac{1}{4} \left[ 1 - \frac{1}{3}\left(
    \phi_{i+1}^{2}+\phi_{i+1}\phi_{i}+\phi_{i}^{2} \right) \right]^{2},
\end{equation}
with the temporal indices equal to $n$. Of course, in the limit $h \rightarrow 0$ both
lattice potentials are equivalent to the continuum one (\ref{phi4pot}). Notice that the
resulting equation of motion involves a field derivative of $V_{\text{TDS}}(\phi_i)$ which is
somewhat intricate \cite{speight2}:

\begin{eqnarray} \label{tdspotderiv}
  V_{\text{TDS}}'(\phi_i) & = & -\frac{1}{6} [ (2\phi_i + \phi_{i-1})(1-\frac{1}{3}(\phi_i^2+\phi_i \phi_{i-1} + \phi_{i-1}^2)) \nonumber \\
    & + & (2\phi_i + \phi_{i+1})(1-\frac{1}{3}(\phi_i^2 + \phi_i \phi_{i+1} + \phi_{i+1}^2)) ]
\end{eqnarray}
This should be contrasted to the simpler expression from a standard discretization:

\begin{equation} \label{potderiv}
  V'(\phi_i) = \phi_i^3 - \phi_i
\end{equation}

  However, the main feature of the TDS potential is that it makes it possible to construct
directly from the lattice theory the Bogomol'nyi lower bound on the kink energy:

\begin{equation} \label{eq_bound}
  0 \leq E_{P} + \frac{h}{\sqrt{2}} \sum_{i} \Delta \left( \frac{1}{3}\phi_i^3 - \phi_i \right) = E_{P} - \frac{2\sqrt{2}}{3},
\end{equation}
with $\Delta$ the forward difference operator $\Delta f_i = (f_{i+1}-f_{i})/h$ and
$E_P$ the lattice potential energy given by:

\begin{equation} \label{lat_potenergy}
  E_P = h \sum_{i} \left[ \frac{1}{2}\left( \Delta \phi_i \right)^2 + V_{\text{TDS}}(\phi_i) \right].
\end{equation}
The kinks are static solutions of the field equations with energy equal to the above lower
limit, namely $2\sqrt{2}/3$.

  The first obvious thing to be done is to reproduce the main numerical results of
\cite{speight2} in order to check our TDS implementation. This is shown in Fig.
\ref{FIG:reprod} for the case of a spatial discretization $h=1.2$ and initial velocity $v=0.6$.
This result is remarkable, and it is tempting to use this lattice in other dynamical
kink problems, as suggested in \cite{speight1}.

\begin{figure}
  \centerline{ \epsfig{figure=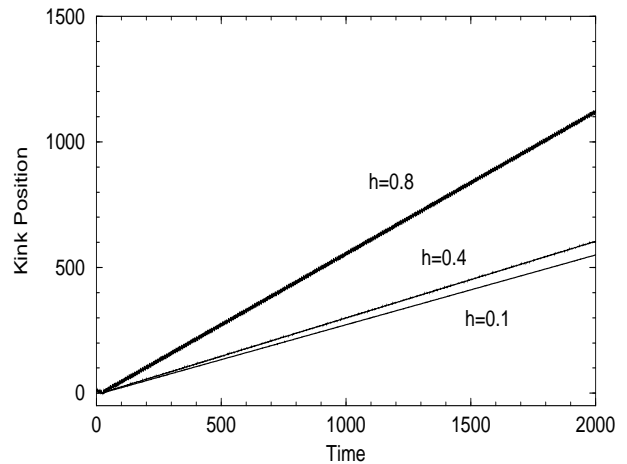,width=230pt,height=180pt} }
  \caption[caption]{
    Kink position after a collision with an antikink (the collision occurred approximately
    at $t\approx23$). All the three curves were obtained from a TDS lattice. The most
    accurate one is, of course, for $h=0.1$, where a standard discretization gives the same plot.
    The final kink velocity, as obtained by taking the slope of the curves for $t>100$,
    are $0.2777$, $0.3049$ and $0.5654$ for $h=0.1$, $0.4$ and $0.8$, respectively.
  }
  \label{FIG:kinkpos}
\end{figure}

  Since we have restricted our study to kink-antikink problems (including bound states, such
as breathers), we benefit from the symmetry of system [$\phi(x,t)=\phi(-x,t)$] so that we
need to integrate only half the original lattice. We have adopted the ``adiabatic damping
method'' of Gleiser and Sornborger \cite{adiabdamp} for the study of kink-antikink bound
states, and a moving boundary with constant velocity equal to that of light (i.e., with $v=1$
in our dimensionless units) for the study of kink-antikink scattering. These two methods are
very effective from the computational point of view, in that they avoid the reflection of
practically all the outgoing radiation by the boundaries without the need of using huge
lattices.

\begin{figure}
  \centerline{ \epsfig{figure=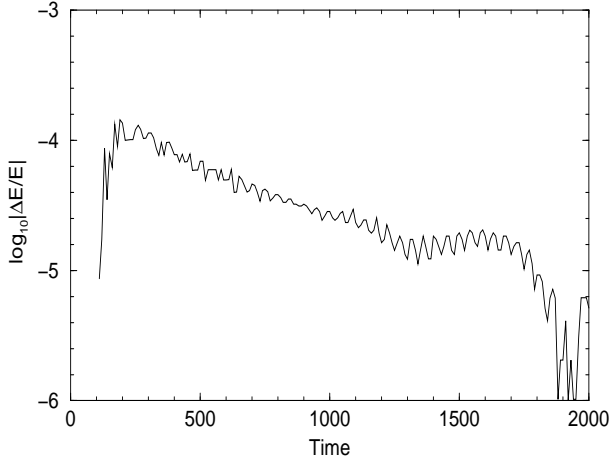,width=230pt,height=180pt} }
  \caption[caption]{
    Relative energy error for the adiabatic damping method of Ref. \cite{adiabdamp} in
    comparison to an outgoing boundary for our breather problem. The data for $t\lesssim100$
    is not shown since $\Delta E=0$ for this case. The parameters, in the notation of the above
    reference, are $k=0.005$, $\rho_{0}=150$ and $L=250$, where $2L$ is the physical size of the
    lattice.
  }
  \label{FIG:logE}
\end{figure}

  By using as a guideline the very precise work of Campbell et al. \cite{campbell}, we have
investigated the behavior of kink-antikink scattering above the critical velocity
$v_c=0.2598...$, where there are no resulting bound states. In Fig. \ref{FIG:kinkpos} we show a
particular sample of our simulations, for the case of an initial kink velocity $v=0.4$
(similar results are obtained for different $v$). The final kink velocity as calculated in the
above mentioned work for this same initial velocity is
$v_{f}\approx0.28$, where the authors have adopted a very fine lattice, namely $h=0.01$.
Comparing this result with ours, we notice that our simulations are quite accurate even
for $h=0.1$ (see notes in Fig. \ref{FIG:kinkpos}). From this figure we also clearly see
that the TDS lattice is accurate only when $h\approx0.1$, where a standard discretization
also works very well (it is not shown in this figure since it coincides with the TDS curve).
Therefore, even though the TDS lattice has achieved a very good constant kink velocity in coarse
lattices, it overestimated the final kink velocity {\em during} the kink-antikink interaction
(notice from Fig. \ref{FIG:kinkpos} that the kink velocity is practically constant for $t>23$).
We suspect that this fact is related to the breaking of the Bogomol'nyi bound, since, during this
interaction, the field configuration loses its static property which is required by the
Bogomol'nyi argument.

  Another application of interest within the context of kink dynamics is the formation
of kink-antikink bound states, resulting in the so-called breathers.
Perhaps the most popular breather is the one in the sine-Gordon theory \cite{dodd}, being
mainly characterized by an exact periodic oscillation of the field such that the resulting
coherent structure is stable, i.e., it retains the energy localization even when
$t\rightarrow\infty$. For the $\phi^4$ theory in one spatial dimension, however, it is today
known that no such stable structure exists \cite{segur}. However, an extremely long-lived
structure very similar to a legitimate breather exists, and has been a topic of active
research not only in one \cite{geicke}, but also in two and three spatial dimensions
\cite{adiabdamp,oscillons}.

\begin{figure}
  \centerline{ \epsfig{figure=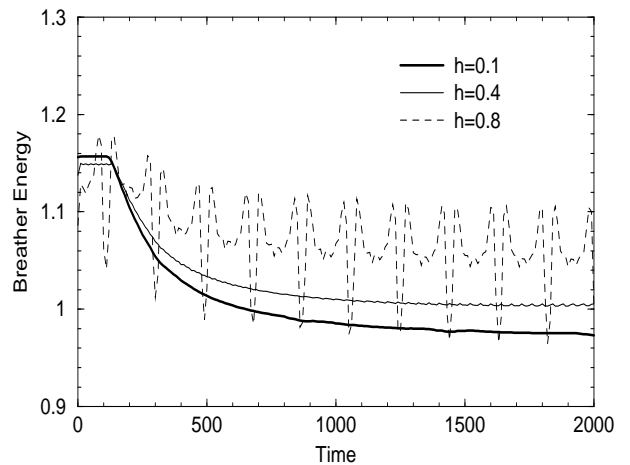,width=230pt,height=180pt} }
  \caption[caption]{
    Breather energy for three different lattice spacing $h$ in a TDS system. Again, the
    $h=0.1$ case coincides with the result of a standard discretization.
  }
  \label{FIG:breathE}
\end{figure}

  Using as a reference the work of Geicke on the logarithmic decay of $\phi^4$ breathers with
energy $E \lesssim 1$ \cite{geicke}, we have also numerically investigated the effectiveness
of the TDS approach and, as already anticipated above, the results are disappointing.
The breather, as obtained in the above work, arises from a dynamical kink-antikink interaction,
with the kinks initially at rest and at a mutual distance equal to $1.6$.
Again, and as a first step, we have checked our numerical routines by comparing our results
with the ones available in the reference work. The main observable adopted here is the energy
within a certain region around $x=0$, i.e., the effective breather energy. By observing
the same region as Geicke, namely $-b \leq x \leq b$ with $b=100$, and integrating up to
$t=2000$, the resulting energy was $E=0.9732$, whereas Geicke's was $E=0.9812$.\footnote{
  \hsize\textwidth\columnwidth\hsize
  Geicke reported two values of the energy: one corresponding to the configuration
  where $\phi(0,t)$ assumed a maximum and the other when $\phi(0,t)$ assumed a minimum. The
  difference, however, is quite small ($\sim0.004$), so here we quote only the smaller one.
} We suspect that this difference is related to the presence of some dim radiation reflected
by the boundaries of Geicke's lattice [which, although alleviated by his boundary conditions,
is still evident from Table I in his work, for which a greater value of the lattice size $a$
would probably follow the observed tendency to decrease $E(t=2000)$ for fixed $b=100$].
As already stated previously, we have adopted a very efficient damping method to avoid this
sort of radiation, and have found a set of parameters for which the method gives a very
accurate result (better than one part in $10^4$, see Fig. \ref{FIG:logE}) in comparison to an
outgoing boundary (which never reflects the radiation, see \cite{adiabdamp}).

  In Fig. \ref{FIG:breathE} we present another sample of our simulations, now for the
breather energy in a TDS lattice for three values of $h$.
We feel that this plot is sufficient to show that a topological lattice gives very
different results even for relatively fine lattices, such as $h=0.4$. For more coarse
lattices, such as $h=0.8$, the results start to differ not only in the magnitude of the
energy, but also in its profile. Of course, for larger still spatial discretizations
the error grows without bounds and therefore the TDS approach is definitely not suitable
for such problems. Again, we believe that this error is due to the breaking of the
Bogomol'nyi bound since, although the kink and the antikink are static solutions of the
field equations separately, the same is far from being true for their bound state, i.e.,
for the breather case.

  We conclude this report by remarking that for all the problems tackled here where the
TDS approach is accurate, i.e. where $h\approx0.1$, a standard discretization is not only
simpler, but also computationally more efficient since it involves fewer
operations (compare Eqs. (\ref{tdspotderiv}) and (\ref{potderiv}), for example).
We cannot disagree with the fact that such topological lattices are indeed a
remarkable discretization for the study of {\em free} kinks on a lattice.
However, in the face of the numerical experiments presented here, we can {\em not} agree
that this construction is an efficient and natural choice for numerically simulating
kink {\em dynamics} in nonlinear continuum Klein-Gordon models \cite{speight2,speight3}.

  A.B.A. would like to thank the Physics Department at UFC for the kind hospitality and
computational resources provided during the preparation of this work. C.A.S.A. was
supported in part by the Brazilian agency CNPq.

\end{multicols}


\end{document}